\documentclass[pra,aps,twocolumn,nopacs,superscriptaddress,nofootinbib,reprint]{revtex4}
\usepackage[T1]{fontenc}
\usepackage[latin1]{inputenc}
\usepackage{lmodern}
\usepackage{graphicx}
\usepackage{dcolumn}
\usepackage{bm}
\usepackage{amsmath}
\usepackage{amssymb}
\usepackage{pst-all}
\usepackage{psfrag}
\usepackage{units}
\usepackage{gensymb}
\usepackage{color}
\usepackage[colorlinks=true, allcolors=blue]{hyperref}

\def\ii{{\rm i}}  \def\ee{{\rm e}}
\def\rb{{\bf r}}  \def\Rb{{\bf R}}    \def\vb{{\bf v}} 
      
  \def\kpar{k_\parallel}  \def\kparb{{\bf k}_\parallel}
\def\Qb{{\bf Q}}     \def\0b{{\bf 0}}
\def\me{m_{\rm e}}  
\def\rp{r_{\rm p}}  
\def\vF{v_{\rm F}}  \def\kF{{k_{\rm F}}}  \def\EF{{E_{\rm F}}}
        \def\wp{\omega_{\rm p}}
\def\EELS{{\rm EELS}}  \def\ext{{\rm ext}}          \def\ind{{\rm ind}}
\def\en{\varepsilon}   \def\ws{\omega_{\rm s}}      \def\uupsilon{\nu}
\def\dm{d}             \def\epsm{\epsilon}  \def\epsb{\epsilon_{\rm b}}

\begin{document}
\title{Theory of EELS in atomically thin metallic films}
\author{A.~Rodr\'{\i}guez~Echarri}
%\thanks{These two authors contributed equally.}
\affiliation{ICFO-Institut de Ciencies Fotoniques, The Barcelona Institute of Science and Technology, 08860 Castelldefels (Barcelona), Spain}
\author{Enok Johannes Haahr Skj{\o}lstrup}
%\thanks{These two authors contributed equally.}
\affiliation{Department of Materials and Production, Aalborg University, Skjernvej 4A, DK-9220 Aalborg East, Denmark}
\author{Thomas G. Pedersen}
\affiliation{Department of Materials and Production, Aalborg University, Skjernvej 4A, DK-9220 Aalborg East, Denmark}
\author{F.~Javier~Garc\'{\i}a~de~Abajo}
\email[Corresponding author:]{javier.garciadeabajo@nanophotonics.es}
\affiliation{ICFO-Institut de Ciencies Fotoniques, The Barcelona Institute of Science and Technology, 08860 Castelldefels (Barcelona), Spain}
\affiliation{ICREA-Instituci\'o Catalana de Recerca i Estudis Avan\c{c}ats, Passeig Llu\'{\i}s Companys 23, 08010 Barcelona, Spain}

\date{\today}

\begin{abstract}
We study strongly confined plasmons in ultrathin gold and silver films by simulating electron energy-loss spectroscopy (EELS). Plasmon dispersion relations are directly retrieved from the energy- and momentum-resolved loss probability under normal incidence conditions, whereas they can also be inferred for aloof parallel beam trajectories from the evolution of the plasmon features in the resulting loss spectra as we vary the impinging electron energy. We find good agreement between nonlocal quantum-mechanical simulations based on the random-phase approximation and a local classical dielectric description for silver films of different thicknesses down to a few atomic layers. We further observe only a minor dependence of quantum simulations for these films on the confining out-of-plane electron potential when comparing density-functional theory within the jellium model with a phenomenological experimentally-fitted potential incorporating atomic layer periodicity and in-plane parabolic bands of energy-dependent effective mass. The latter shows also a small dependence on the crystallographic orientation of silver films, while the unphysical assumption of energy-independent electron mass leads to spurious features in the predicted spectra. Interestingly, we find electron band effects to be more relevant in gold films, giving rise to blue shifts when compared to classical or jellium model simulations. In contrast to the strong nonlocal effects found in few-nanometer metal nanoparticles, our study reveals that a local classical description provides excellent quantitative results in both plasmon strength and dispersion when compared to quantum-mechanical simulations down to silver films consisting of only a few atomic layers, thus emphasizing the in-plane nearly-free conduction-electron motion associated with plasmons in these structures.
\end{abstract}

\maketitle

{\bf Physics Subject Headings:} EELS; Surface plasmons; Thin films; Quantum-well states; Nanophotonics; Nonlocal effects.

% ----------------------------------------------------------------------------
\section{Introduction}

Surface plasmons --the collective electron oscillations at material surfaces and interfaces-- provide the means to concentrate and amplify the intensity of externally applied light down to nanoscale regions \cite{BS03,XBK99}, where they interact strongly with molecules and nanostructures, thus becoming a powerful asset in novel applications \cite{P08_2} such as biosensing \cite{XBK99,AHL08,paper256}, photocatalysis \cite{SLL12_2,C14}, energy harvesting \cite{AP10,CTF14}, and nonlinear optics \cite{DN07,B08_3,DSK08,PN08}. 

Surface plasmons were first identified using electron energy-loss spectroscopy (EELS), starting with the prediction \cite{R1957} and subsequent measurement of associated loss features in electrons scattered under grazing incidence from Al \cite{PS1959}, Na and K \cite{TPF1989,RYB95}, and Ag \cite{D1967,RYB95} surfaces. The main characteristics of surface plasmons in noble and simple metals were successfully explained using time-dependent density-functional theory (TD-DFT) \cite{L93} within the jellium model \cite{LK1970_2,LK1973}, while inclusion of electron band effects were required for other metals \cite{PSC07}. Interestingly, multipole surface plasmons were predicted as additional resonances originating in the smooth electron density profile across metal-dielectric interfaces \cite{DH1988,EYQ1975,PSC07}, and subsequently found in experiments performed on simple metals such as K and Na \cite{TPL90}, but concluded to be too weak to be observed in Al \cite{TPL90} and Ag \cite{LS95}. These studies focused on the relatively high-energy plasmons supported by planar surfaces in the short-wavelength regime. However, plasmons can hybridize with light forming surface-plasmon polaritons (SPPs) in planar surfaces, which become light-like modes at low energies, thus loosing confinement, as they are characterized by in-plane wavelengths slightly smaller than those of light and long-range penetration into the dielectric material or empty space outside the metal \cite{R1988,M07,BDE03}.

Highly confined plasmons can also be achieved in sharp metallic tips and closely spaced metal surfaces \cite{paper156}, where strong redshifts are produced due to the attractive Coulomb interaction between neighboring non-coplanar interfaces. This effect, which depends dramatically on surface morphology, can also be observed in planar systems such as ultrathin noble metal films \cite{E1969,paper335} and narrow metal-dielectric-metal waveguides \cite{E1969,DSK08,paper329}. More precisely, hybridization takes place in metal films between the plasmons supported by their two interfaces, giving rise to bonding and antibonding dispersion branches that were first revealed also through EELS is self-standing aluminum foils \cite{VS1973}; in ultrathin films of only a few atomic layers in thickness, the antibonding plasmon dispersion is pushed close to the light line, whereas the bonding plasmon becomes strongly confined (reaching the quasistatic limit \cite{E1969,paper335}), as experimentally corroborated through angle-resolved low-energy EELS in few-monolayer Ag films \cite{MRH99} and monolayer (ML) DySi$_2$ \cite{RNP08}, as well as in laterally confined wires formed by In \cite{CKH10} and silicide \cite{RTP10}, and even in monoatomic Au chains grown on Si(557) surfaces \cite{NYI06}. Additionally, graphene has been shown to support long-lived mid-infrared and terahertz plasmons \cite{NMS18} that can be tuned electrically \cite{FRA12,paper196} and confined vertically down to few nanometers when placed in close proximity to a planar metal surface \cite{LGA17,AND18}. While most graphene plasmon studies have been performed using far- and near-field optics setups \cite{JGH11,GPN12,paper283}, low-energy EELS has also revealed their dispersion relation in extended films \cite{LWE08,LW10}. Here, we focus instead on visible and near-infrared plasmons supported by atomically thin metal films, which have been recently demonstrated in crystalline Ag layers \cite{paper335}, where they also experience strong spatial confinement.

In this paper, we investigate plasmons in atomically thin noble metal films by theoretically studying EELS for electron beams either traversing them or moving parallel outside their surface. We provide quantum-mechanical simulations based on the random-phase approximation (RPA), which are found to be in excellent agreement with classical dielectric theory based on the use of frequency-dependent dielectric functions for both Ag and Au films of small thickness down to a few atomic layers. This result is in stark contrast to the strong nonlocal effects observed in metal nanoparticles of similar or even larger diameter \cite{KV95,SKD12}, a result that we attribute to the predominance of in-plane electron motion associated with the low-energy plasmons of thin films, unlike the combination of in- and out-of-plane motion in higher energy SPPs.

% ----------------------------------------------------------------------------
\section{Theoretical formalism}
\label{tf}

We present the elements needed to calculate EELS probabilities in the nonretarded approximation using the linear response susceptibility to represent the metallic thin film. The latter is obtained in the RPA, starting from the one-electron wave functions of the system, which are organized as vertical quantum-well (QW) states, discretized by confinement along the out-of-plane direction and exhibiting quasi-free motion along the plane of the film. We further specify the EELS probability for electron trajectories either parallel or perpendicular with respect to the metal surfaces.

% ----------------------------------------------------------------------------
\subsection{Calculation of EELS probabilities from the susceptibility in the nonretarded limit}

The loss probability $\Gamma^\EELS(\omega)$ measured through EELS in electron microscopes must be normalized in such a way that $\int_0^\infty d\omega\,\hbar\omega\,\Gamma^\EELS(\omega)$ gives the average energy loss experienced by the electrons. Taking the latter to follow a straight-line trajectory with constant velocity vector $\vb$ parallel to the $z$ axis and impact parameter $\Rb_0=(x_0,y_0)$, we can write \cite{paper149}
\begin{align}
\Gamma^\EELS(\omega)=\frac{e}{\pi\hbar\omega}\int dz\,{\rm Re}\left\{E_z^\ind(\Rb_0,z,\omega)\,\ee^{-\ii\omega z/v}\right\}
\label{eq:EELS}
\end{align}
as the integral along the electron trajectory of the frequency-resolved self-induced field $E_z^\ind(\rb,\omega)=\int dt\,E_z(\rb,t)\ee^{\ii\omega t} $, which can be in turn calculated by solving the classical Maxwell equations with the electron point charge acting as an external source in the presence of the sample. This equation is rigorously valid within the approximations of linear response and nonrecoil (i.e., small energy loss $\hbar\omega$ compared with the electron kinetic energy $E_0$).

In the present study, we consider relatively small electron velocities $v\ll c$ and films of small thickness compared with the involved optical wavelengths. This allows us to work in the quasistatic limit and write the field $E_z^\ind(\rb,\omega)=-\partial_z\phi^\ind(\rb,\omega)$ as the gradient of a scalar potential, so Eq.\ (\ref{eq:EELS}) can be integrated by parts to yield
\begin{align}
\Gamma^\EELS(\omega)=\frac{e}{\pi\hbar v}\int dz \, {\rm Im}\left\{\phi^\ind(\Rb_0,z,\omega)\,\ee^{-\ii\omega z/v}\right\}.
\label{eq:EELSNR}
\end{align}
We can now express the induced potential in terms of the induced charge as
\begin{align}
\phi^\ind(\rb,\omega)=\int d^3\rb'\,\uupsilon(\rb,\rb')\,\rho^\ind(\rb',\omega),
\label{rhoind}
\end{align}
where $\uupsilon(\rb,\rb')$ is the Coulomb interaction between point charges located at positions $\rb$ and $\rb'$. Likewise, we write the induced charge as $\rho^\ind(\rb,\omega)=\int d^3\rb'\,\chi(\rb,\rb',\omega)\phi^\ext(\rb',\omega)$, where $\chi(\rb,\rb',\omega)$ is the linear susceptibility, $\phi^\ext(\rb',\omega)=\int d^3\rb'\,\uupsilon(\rb,\rb')\,\rho^\ext(\rb',\omega)$ is the external electric potential generated by the electron charge density $\rho^\ext(\rb,\omega)=-e\int dt\,\delta(\rb-\Rb_0-\vb t)\,\ee^{\ii\omega t}=(-e/v)\delta(\Rb-\Rb_0)\,\ee^{\ii\omega z/v}$, and we use the notation $\rb=(\Rb,z)$ with $\Rb=(x,y)$.

In free space one has $\uupsilon(\rb,\rb')=\uupsilon_0(\rb-\rb')=1/|\rb-\rb'|$, but we are interested in retaining a general spatial dependence of $\uupsilon(\rb,\rb')$ in order to describe the polarization background produced in the film by interaction with everything else other than conduction electrons (see below). Combining these elements with Eq.\ (\ref{eq:EELSNR}), we find the loss probability
\begin{align}
\Gamma^\EELS(\omega)=\frac{e^2}{\pi\hbar v^2 }&\int d^3\rb \int d^3\rb' \,
w^*(\rb)\,w(\rb') \label{eq:EELS_general} \\
&\times{\rm Im}\left\{-\chi(\rb,\rb',\omega)\right\}, \nonumber
\end{align}
where
\begin{align}
w(\rb)=\int dz' \uupsilon(\rb,\Rb_0,z')\,\ee^{\ii\omega z'/v} \label{wRz}
\end{align}
is the {\it external} potential created by the electron and we have made use of the reciprocity property $\chi(\rb,\rb',\omega)=\chi(\rb',\rb,\omega)$ to extract the complex factors $w$ outside the imaginary part. Next, we apply this expression to calculate EELS probabilities from the RPA susceptibility. But first, for completeness, we note that the integral in Eq.\ (\ref{wRz}) can be performed analytically for the bare Coulomb interaction \cite{GR1980} yielding
\begin{align}
w(\rb)=2K_0(\omega|\Rb-\Rb_0|/v)\,\ee^{\ii\omega z/v}, \nonumber
\end{align}
where $K_0$ is a modified Bessel function \cite{GR1980}, thus allowing us to write
\begin{align}
\Gamma^\EELS(\omega)=&\frac{4e^2}{\pi\hbar v^2 }\int d^3\rb \int d^3\rb' \, 
\cos\left[\frac{\omega}{v} (z'-z)\right] \nonumber \\
&\times K_0\left(\frac{\omega}{v}|\Rb-\Rb_0|\right)
K_0\left(\frac{\omega}{v}|\Rb'-\Rb_0|\right) \nonumber\\
&\times{\rm Im}\left\{-\chi(\rb,\rb',\omega)\right\} \nonumber
\end{align}
for the loss probability, which we can directly apply to systems in which any background polarization is already contained in $\chi$, or when $\uupsilon$ is well described by the bare Coulomb interaction (e.g., in simple metals).

% ----------------------------------------------------------------------------
\subsection{RPA susceptibility of thin metal films}
\label{Rso}

We follow the same formalism as in Ref.\ \cite{paper329}, which is extended here to account for an energy-dependence of the in-plane electron effective mass. One starts by writing $\chi(\rb,\rb',\omega)$ in terms of the non-interacting susceptibility $\chi^0(\rb,\rb',\omega)$ through $\chi=\chi^0\cdot(\mathcal{I}-\uupsilon\cdot\chi^0)^{-1}$, where we use matrix notation with spatial coordinates $\rb$ and $\rb'$ acting as matrix indices, so that matrix multiplication involves integration over $\rb$, and $\mathcal{I}(\rb,\rb')=\delta(\rb-\rb')$. We further adopt the RPA by calculating $\chi^0$ as \cite{HL1970,paper329}
\begin{align}
\chi^0(\rb,\rb',\omega)=\frac{2e^2}{\hbar} \sum_{ii'} \left(f_{i'}-f_i\right) \frac{\psi_i(\rb)\psi_i^*(\rb')\psi_{i'}^*(\rb)\psi_{i'}(\rb')}{\omega+\ii\gamma-(\en_i-\en_{i'})} \label{eq:chi_0r1}
\end{align}
from the one-electron wave functions $\psi_i$ of energies $\hbar\en_i$ and Fermi-Dirac occupation numbers $f_i$. Here, the factor of 2 accounts for spin degeneracy and $\gamma$ is a phenomenological damping rate.

We describe metal films assuming translational invariance along the in-plane directions and parabolic electron dispersion with different effective mass $m_j^*$ for each vertical QW band $j$. This allows us to write the electron wave functions as \cite{K1986} $\psi_i(\rb)=\varphi_j(z)\ee^{\ii\kparb\cdot\Rb}/\sqrt{A}$, where $\kparb$ is the 2D in-plane wave vector, $A$ is the quantization area, and the state index is multiplexed as $i\rightarrow(j,\kparb)$. Likewise, the electron energy can be separated as $\hbar\en_{j,\kpar}=\hbar\en_j^\perp+\hbar^2\kpar^2/2m^*_j$, where $\hbar\en_j^\perp$ is the out-of-plane energy that signals the QW band bottom. Inserting these expressions into Eq.\ (\ref{eq:chi_0r1}) and making the customary substitution $\sum_i\rightarrow A\sum_j\int d^2\kparb/(2\pi)^2$ for the state sums, we find \cite{MS01}
\begin{align}
\chi(\rb,\rb',\omega)=\int \frac{d^2\Qb}{(2\pi)^2}\,\chi(Q,z,z',\omega)\,\ee^{\ii\Qb\cdot(\Rb-\Rb')},
\label{chiQzz}
\end{align}
which directly reflects the in-plane homogeneity of the film. We can now work in $\Qb$ space, where Eq.\ (\ref{eq:chi_0r1}) reduces, using the above assumptions for the wave functions, to
\begin{align}
&\chi^0(Q,z,z',\omega)  \label{eq:chi0_Q1}\\
&= \frac{2e^2}{\hbar} \sum_{jj'} \chi_{jj'}(Q,\omega) \varphi_{j}(z)\varphi^*_{j}(z')\varphi^*_{j'}(z)\varphi_{j'}(z') \nonumber
\end{align}
where
\begin{align}
&\chi_{jj'}(Q,\omega)=\int \frac{d^2\kparb}{(2\pi)^2}\,\left(f_{j',|\kparb-\Qb|}-f_{j,\kpar}\right)
\label{eq:I_jj2}\\
&\times\frac{1}{\omega +\ii \gamma- \left[\en^\perp_j-\en^\perp_{j'}+\frac{\hbar}{2}\left(\kpar^2/m^*_j-|\kparb-\Qb|^2/m^*_{j'}\right) \right]}, \nonumber
\end{align}
which only depends on the modulus of $\Qb$ due to the in-plane band isotropy. We evaluate the integral in Eq.\ (\ref{eq:I_jj2}) assuming zero temperature [i.e., $f_{j,\kpar}=\theta(\EF-\hbar\en_{j,\kpar})$, where $\EF$ is the Fermi energy] and taking $\Qb=(Q,0)$ without loss of generality.
% This allows us to perform the $k_y$ integral analytically, while the remaining $k_x$ integral is carried out numerically.

Incidentally, simple manipulations of the above expressions reveal a dependence on frequency and damping through $(\omega+\ii\gamma)^2$ that is maintained in the local limit ($Q\rightarrow0$), in contrast to $\omega(\omega+\ii\gamma)$ in the Drude model. The RPA formalism thus produces spectral features with roughly twice the width of the Drude model in the local limit. This problem (along with a more involved issue related to local conservation of electron number for finite attenuation) can be solved through a phenomenological prescription proposed by Mermin \cite{M1970}, which unfortunately becomes rather involved when applied to the present systems. As a practical and reasonably accurate solution, we proceed instead by setting $\gamma=\gamma^{\rm exp}/2$ in the above expressions (i.e., half the experimental damping rate, see Appendix\ \ref{bsi}).

\begin{table*}
\centering
\begin{tabular}{c|cccccc}  \hline
Material & $a (\rm eV^{-1}$) & $b$ & $m^*$(SS)$/\me$ & $m_0/\me$& $n_{\rm eff}/n_0$ & $\EF$ (eV) \\ \hline
Ag(111) &  -0.1549  & -0.5446 & 0.40 \cite{RNS01} & 0.25 \cite{SPC05} & 0.8381 & -4.63 \cite{PMM95} \\
Ag(100) &  -0.0817  & 0.2116  & -                 & 0.40 \cite{GPC03} & 0.8710 & -4.43 \cite{CSE99} \\
Au(111) &  -0.1660  & -0.8937 & 0.26 \cite{RNS01} & 0.26 \cite{SPC05} & 0.9443 & -5.50 \cite{PMM95} \\ \hline
\end{tabular}
\caption{Parameters used to describe the parabolic dispersion of quantum wells (QWs) in Ag(111), Ag(100), and Au(111) films. We take the effective mass of each QW $j$ to linearly vary as $m^*_j/\me=a\hbar\en_j^\perp+b$ with band-bottom energy $\hbar\en_j^\perp$, where the parameters $a$ and $b$ are taken to match $m_0$ at the highest occupied QW (below the SS) in the semi-infinite surface and $m^*=\me$ at the bottom of the conduction band. The effective electron density $n_{\rm eff}$, given here relative the bulk conduction electron density $n_0$, is required to fit the experimentally observed Fermi energy $\EF$ and SS energy.}
\label{Table1}
\end{table*}

We obtain the out-of-plane wave functions $\varphi_{j}(z)$ as the eigenstates of the 1D Hamiltonian $-(\hbar^2/2\me)\partial_{zz}+V(z)$, using the free-electron mass $\me$ for the transversal kinetic term and two different models for the confining potential $V(z)$: ({\it i}) the self-consistent solution in the jellium (JEL) approximation within density-functional theory (DFT) \cite{LK1970_2,LK1973}; and ({\it ii}) a phenomenological atomic-layer potential (ALP) that incorporates out-of-plane bulk atomic-layer corrugation and a surface density profile with parameters fitted to reproduce relevant experimental band structure features, such as affinity, surface state energy, and projected bulk band gap, which depend on material and crystal orientation as compiled in Ref.\ \cite{CSE99}.

The JEL model corresponds to the self-consistent DFT solution for a thin slab of background potential and energy-independent effective mass $m_j^*=\me$ \cite{LK1970_2,LK1973}, computed here through an implementation discussed elsewhere \cite{SSP19}.

In the ALP model we fit $m_j^*$ to experimental data (see Table\ \ref{Table1}) and consider an effective electron density $n_{\rm eff}$. Upon integration over the density of states of the parabolic QW bands, we can then write the Fermi energy of a $N$-layer film as
\begin{align}
\EF=\left(\sum_{j=1}^Mm^*_j\right)^{-1}\left(n_{\rm eff} a_s N \hbar^2\pi+\sum_{j=1}^M m^*_j\hbar\en_j^\perp\right), \label{eq:EF}
\end{align}
where $j=M$ is the highest partially populated QW band (i.e., $\en_M^\perp<\EF/\hbar<\en_{M+1}^\perp$) and $a_s$ is the atomic interlayer spacing (i.e., the film thickness is $\dm=Na_s$, with $a_s=0.236$\,nm for Ag(111) and Au(111), and $a_s=0.205$\,nm for Ag(100)). This expression reduces to a similar one in Ref.\ \cite{paper329} when $m^*_j$ is independent of $j$. We adjust $n_{\rm eff}$ for each type of metal surface in such a way that Eq.\ (\ref{eq:EF}) gives the experimental bulk values of $\EF$ listed in Table \ref{Table1}. Incidentally, although the effective mass of surface states also varies with energy \cite{PRH05,K13}, we take it as constant because of the lack of data for ultrathin Au and Ag films; this should be a reasonable approximation for films consisting of $N\geq$5 layers, where the surface state energy is already close to the semi-infinite surface level.

Conduction electrons interact through the bare Coulomb potential in simple metals, which in $\Qb$ space reduces to $\uupsilon(Q,z,z')=(2\pi/Q)\ee^{-Q|z-z'|}$. However, polarization of inner electronic bands plays a major role in the dielectric response of Ag and Au. We describe this effect by modifying $\uupsilon(Q,z,z')$ in order to account for the interaction between point charges in the presence of a dielectric slab of local background permittivity fitted to experimental data \cite{JC1972} after subtracting a Drude term representing conduction electrons (see Appendix\ \ref{bsi}). We thus adopt the local response approximation for this contribution originating in localized inner electron states, whereas conduction electrons are treated nonlocally through the above RPA formalism. Similar to Eq.\ (\ref{chiQzz}), translational symmetry in the film allows us to write
\begin{align}
\uupsilon(\rb,\rb')=\int \frac{d^2\Qb}{(2\pi)^2}\,\uupsilon(Q,z,z')\,\ee^{\ii\Qb\cdot(\Rb-\Rb')},
\label{uQzz}
\end{align}
where $\uupsilon(Q,z,z')$ is reproduced for convenience from Ref.\ \cite{paper329} in Appendix\ \ref{bsi}. We note that Eq.\ (\ref{uQzz}) neglects the effect of lateral atomic corrugation in this interaction (i.e., the background permittivity is taken to be homogeneous inside the film). 

Finally, we calculate $\chi(Q,z,z',\omega)$ from the noninteracting susceptibility [Eq.\ (\ref{eq:chi0_Q1})] and the screened interaction by discretizing both of them in real space coordinates ($z,z'$) and numerically performing the linear matrix algebra explained above. We obtain converged results with respect to the number of discretization points and also compared with an expansion in harmonic functions \cite{paper329}.

% ----------------------------------------------------------------------------
\subsection{EELS probability under normal incidence}

Direct insertion of Eqs.\ (\ref{chiQzz}) and (\ref{uQzz}) into Eqs.\ (\ref{eq:EELS_general}) and (\ref{wRz}) leads to the result
\begin{align}
\Gamma^\EELS_\perp(\omega)=\int_0^\infty \,dQ\,\Gamma^\EELS_\perp(Q,\omega) \label{eq:EELS_perp}
\end{align}
with
\begin{align}
&\Gamma^\EELS_\perp(Q,\omega)=\frac{e^2\,Q}{2\pi^2\hbar v^2} \label{eq:EELS_perp_kernel}\\
&\times\int dz\!\!\int dz'\,I_\perp^*(Q,z)I_\perp(Q,z')\;{\rm Im}\left\{-\chi(Q,z,z',\omega)\right\}, \nonumber
\end{align}
where
\begin{align}
I_\perp(Q,z)=\int dz'\,\uupsilon(Q,z,z')\,\ee^{\ii\omega z'/v} \label{Iperp}
\end{align}
contains the external electron potential. For completeness, we note that when $\uupsilon(Q,z,z')$ is the bare Coulomb interaction $(2\pi/Q)\ee^{-Q|z-z'|}$, Eq.\ (\ref{Iperp}) becomes $I_\perp(Q,z)=4\pi\ee^{\ii\omega z/v}/(Q^2+\omega^2/v^2)$, so Eq.\ (\ref{eq:EELS_perp_kernel}) reduces to
\begin{align}
&\Gamma^\EELS_\perp(Q,\omega)=\frac{8e^2}{\hbar v^2 }\,\frac{Q}{(Q^2+\omega^2/v^2)^2} \label{aaaa}\\
&\times\int dz\int dz'\,\cos\left[\omega(z-z')/v\right]\;{\rm Im}\left\{-\chi(Q,z,z',\omega)\right\}, \nonumber
\end{align}
where we have used reciprocity again [i.e., $\chi(Q,z,z',\omega)=\chi(Q,z',z,\omega)$].

In the simulations that we present below, we compare the RPA approach just presented with classical electromagnetic calculations based on the use of a local frequency-dependent dielectric function for the metal. This configuration has been theoretically studied for a long time \cite{K1968_2}, and in particular, we use the analytical expressions derived in a previous publication for an electron normally incident on a dielectric slab \cite{paper052} with the bulk contribution integrated up to a cutoff wave vector $Q=5\,$nm$^{-1}$.

% ----------------------------------------------------------------------------
\subsection{EELS probability in the aloof configuration}

For an electron moving parallel to the film at a distance $z_0$ from the metal surface, it is convenient to make the substitutions $z\rightarrow x$, $\Rb\rightarrow(y,z)$, and $\Rb_0\rightarrow(0,z_0)$ in Eqs.\ (\ref{eq:EELS_general}) and (\ref{wRz}), so combining them with Eqs.\ (\ref{chiQzz}) and (\ref{uQzz}), and retaining $\Rb=(x,y)$ in the latter, we readily obtain
\begin{align}
&\Gamma^\EELS_\parallel(\omega)=\frac{e^2 L}{\pi^2\hbar v^2} \int_0^\infty dQ_y \label{eq:EELS_para}\\
&\times\!\!\int dz\!\int dz'
\uupsilon^*(Q,z,z_0)\uupsilon(Q,z',z_0)
\;{\rm Im}\left\{-\chi(Q,z,z',\omega)\right\}, \nonumber
\end{align}
where $Q=\sqrt{\omega^2/v^2+Q_y^2}$ and $L$ is the electron path length. Again for completeness, when $\uupsilon(Q,z,z')$ is the bare Coulomb interaction, Eq.\ (\ref{eq:EELS_para}) reduces to \begin{align}
&\Gamma^\EELS_\parallel(\omega)=\frac{4e^2 L}{\hbar v^2} \int_0^\infty \frac{dQ_y}{Q^2} \nonumber\\
&\times\int dz \int dz'
\ee^{-Q\left(|z-z_0|+|z'-z_0|\right)}
\;{\rm Im}\left\{-\chi(Q,z,z',\omega)\right\}. \nonumber
\end{align}
The above expressions can be applied to electron impact parameters $z_0$ both inside or outside the metal, but they can be simplified when the beam is not overlapping the conduction electron charge [see Fig.\ \ref{Fig3}(a)], so that $z_0>z,z'$ in the region inside the above integrals in which $\chi(Q,z,z',\omega)$ is nonzero, and therefore, changing the variable of integration from $Q_y$ to $Q$, we can write
\begin{align}
\Gamma^\EELS_\parallel(\omega)=\frac{2e^2 L}{\pi\hbar v^2 }\int_{\omega/v}^\infty \,dQ\,\frac{\ee^{-2Qz_0}}{\sqrt{Q^2-\omega^2/v^2}}
\,{\rm Im}\{\rp(Q,\omega)\}, \label{eq:EELS_par}
\end{align}
where
\begin{align}
\rp(Q,\omega)&=-\frac{Q}{2\pi}\int dz \int dz' \;\uupsilon^*(Q,z,z_0)\uupsilon(Q,z',z_0) \nonumber\\
&\times\ee^{2Qz_0}\;\chi(Q,z,z',\omega) \label{eq:r}
\end{align}
is the Fresnel reflection coefficient of the film for p polarization in the quasistatic limit. Incidentally, Eq.\ (\ref{eq:r}) is independent of the source location $z_0$ when it does not overlap the metal because $\uupsilon(Q,z,z_0)$ then depends on $z_0$ only through a factor $\ee^{-Q z_0}$ (see Appendix\ \ref{bsi}). Equation\ (\ref{eq:EELS_par}), which agrees with previous derivations from classical dielectric theory \cite{paper228}, reveals ${\rm Im}\{\rp(Q,\omega)\}$ as a loss function, which is used below to visualize the surface plasmon dispersion. We also provide results from a local dielectric description based on the textbook solution of the Poisson equation for the reflection coefficient \cite{paper329}
\begin{align}
\rp^{\rm classical}=\frac{(\epsm^2-1)\left(1-\ee^{-2Q\dm}\right)}{(\epsm+1)^2-(\epsm-1)^2\ee^{-2Q\dm}}
\label{eq:r_FP}
\end{align}
for a metal film of thickness $\dm$ and permittivity $\epsm$.

% ----------------------------------------------------------------------------
\section{Results and discussion}
\label{sec:results}

\begin{figure}
\centering
\includegraphics[width=0.45\textwidth]{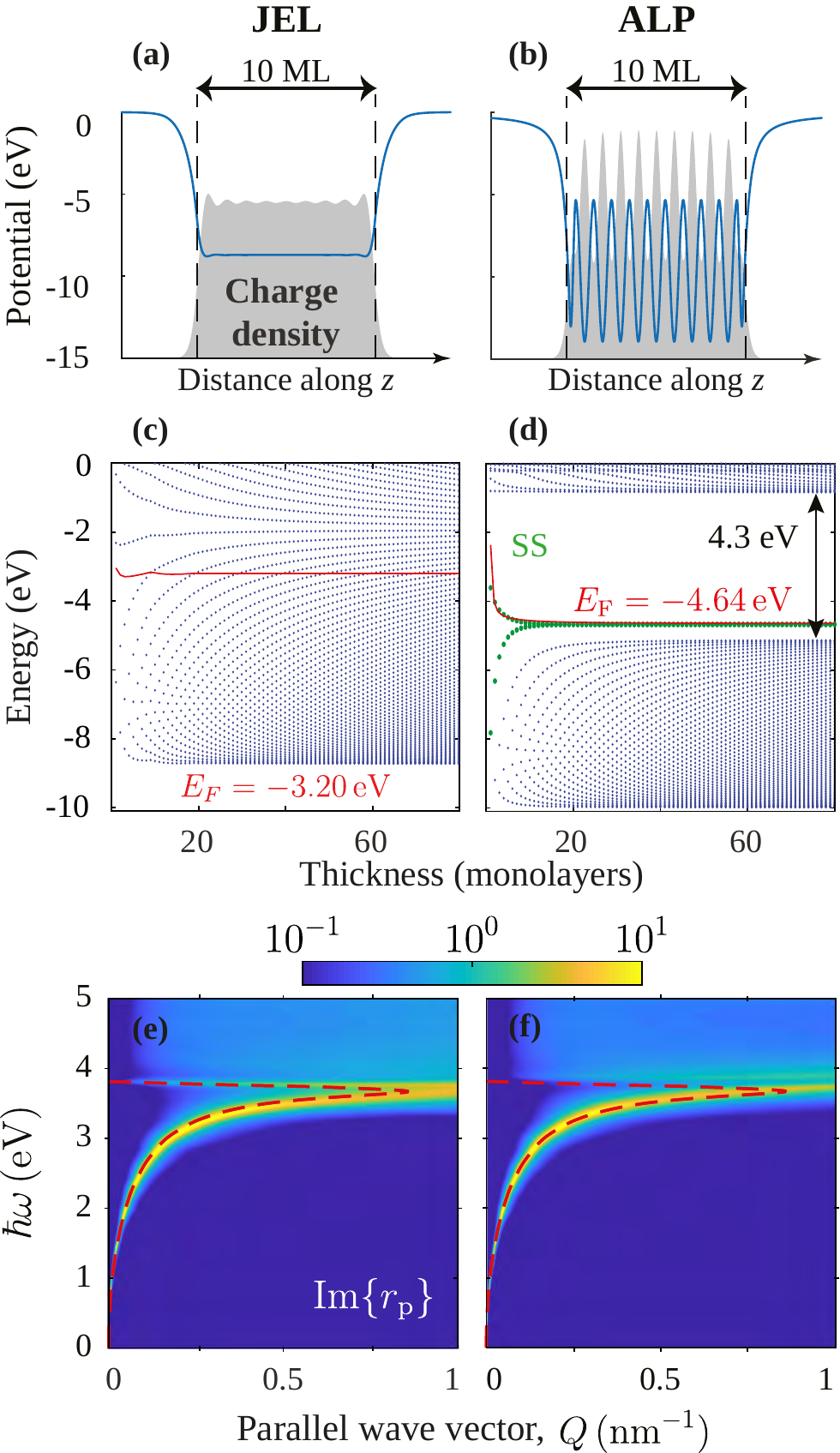}
\caption{RPA description of plasmons in atomically thin Ag(111) films. (a,b) Effective confining potential for conduction electrons across a 10\,ML film. The conduction charge density is shows as shaded areas. (d,c) Electronic energies as a function of film thickness expressed as the number of (111) atomic layers (blue dots). Red curves and green dots represent the Fermi energy and the surface states (SSs). (e,f) Loss function ${\rm Im}\{\rp\}$ calculated in the RPA [color plot, Eq.\ (\ref{eq:r})], compared with the plasmon dispersion relation in the local Drude dielectric model (red curves). Left (right) panels are calculated in the jellium (ALP) model.}
\label{Fig1} 
\end{figure}

We show examples of the two types of confining electron potentials used in our RPA calculations for Ag films in Fig.\ \ref{Fig1}(a,b), along with the resulting conduction electron charge densities. The JEL potential is smooth at the surface and describes electron spill-out and Friedel oscillations \cite{AM1976}. The phenomenological ALP potential further incorporates corrugations due to the atomic planes in the bulk, which result in strong oscillations of the density. The computed electron energies $\hbar\en_j$ (see Sec.\ \ref{Rso}), which correspond to the bottom points of the QW bands (i.e., for vanishing in-plane momentum), are distributed with $N$ of them below the Fermi level in a Ag(111) film consisting of $N$ monolayers [Fig.\ \ref{Fig1}(c,d)]. The band structure quickly evolves toward the semi-infinite surface for a few tens of MLs in both models. Additionally, the ALP potential hosts surface states and a projected bulk gap of energies fitted to experiment \cite{CSE99}. We note that this gap depends on surface orientation: it is present in Ag(111) but absent in Ag(100) at the Fermi level, as revealed by photoemission measurements \cite{GDB198`5} see also Fig.\ \ref{FigS1}(a) in Appendix\ \ref{AFSI}]. Remarkably, despite the important differences in the details of the potentials and electron bands, both models predict a similar plasmon dispersion [Fig.\ \ref{Fig1}(e,f), density plots, obtained from Eq.\ (\ref{eq:r})], which is in excellent agreement with classical theory [Fig.\ \ref{Fig1}(e,f), red curves, obtained from the poles of Eq.\ (\ref{eq:r_FP})]. Incidentally, we observe the response to also converge toward the semi-infinite surface limit for a few tens of atomic layers [see Fig.\ \ref{FigS2} in Appendix\ \ref{AFSI}. Similar good agreement is found in the reflection coefficients of Ag films computed for different thickness with either of these potentials, with a square-barrier potential, or with a model potential constructed by glueing on either film side a jellium DFT potential tabulated for semi-infinite surfaces \cite{LK1970_2} [see Fig.\ \ref{FigS3} in Appendix\ \ref{AFSI]}.

\begin{figure}
\centering
\includegraphics[width=0.40\textwidth]{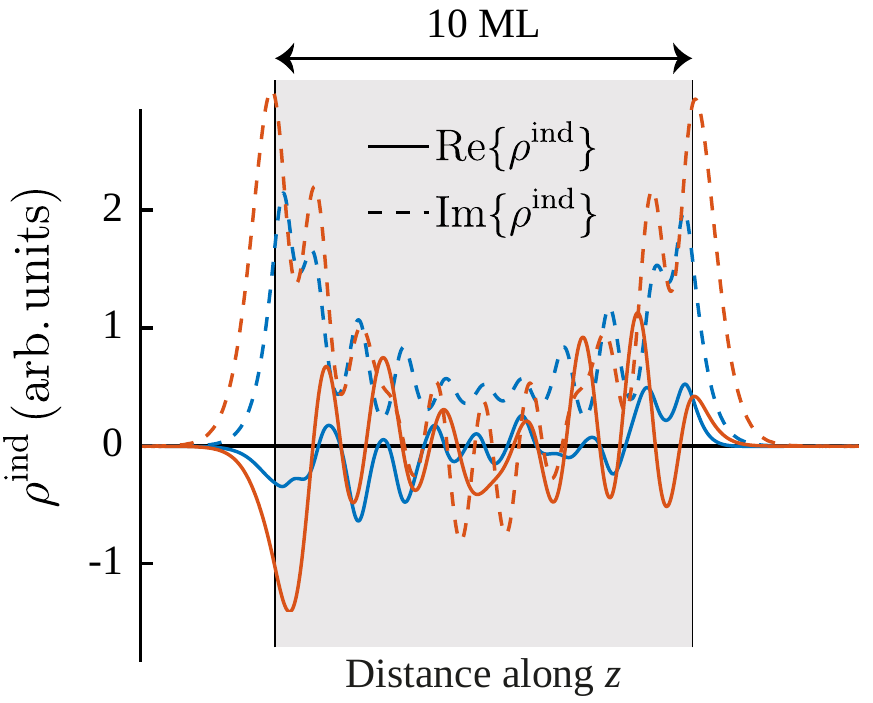}
\caption{Plasmon charge density across a thin 10\,ML Ag(111) film. We plot the real (solid curves) and imaginary (dashed curves) parts of the induced charge density $\rho^\ind$ as calculated in the RPA for excitation by a source placed to the left of the film at the plasmon energies $\hbar\omega=3.54$\,eV and $\hbar\omega=3.47$\,eV corresponding to a parallel wave vector $Q=0.5$ nm$^{-1}$ in the ALP (blue) and JEL (orange) models, respectively.}
\label{Fig2} 
\end{figure}

The transversal distribution of change densities associated with thin film plasmons show a clear resemblance when calculated using the ALP or JEL model potentials, although one can still observe substantial discrepancies between the two of them [see for example Fig.\ \ref{Fig2}, where the ALP model charge appears to be smaller in magnitude]. However, this different behavior hardly reflects in the dispersion relation and plasmon strength [Fig.\ \ref{Fig1}]. Interestingly, the $z$-integrated charge is nonzero, revealing that plasmons involves net charge oscillations along the in-plane directions for finite wave vector $Q$.
%and its imaginary part remains always positive, unlike the JEL model charge

We conclude from these results that it is the effective number of valence electrons participating in the plasmons what determines their main characteristics, irrespective of the details of the electron wave functions and induced charge densities.

\begin{figure}
\centering
\includegraphics[width=0.45\textwidth]{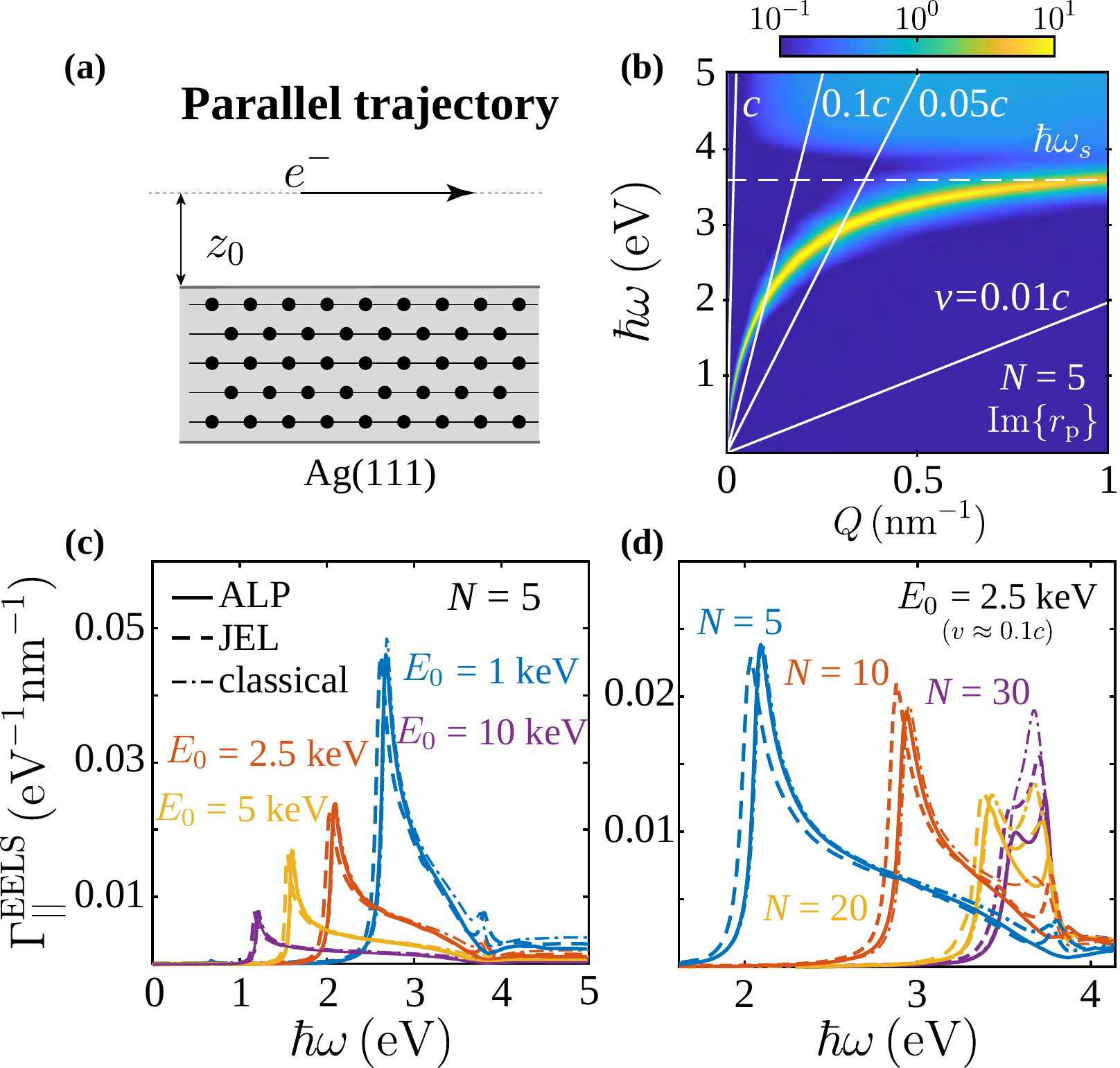}
\caption{Aloof EELS in thin Ag(111) films. (a) Scheme showing an electron moving parallel to a $N=5$\,ML Ag(111) metal film at a distance $z_0$ from its upper surface. (b) Dispersion diagram showing ${\rm Im}\{\rp\}$ calculated in the ALP model for the film shown in (a). White solid lines correspond to $\omega=vQ$ for different velocities $v$, while the dashed horizontal line shows the classical high-$Q$ asymptotic surface-plasmon energy $\hbar\ws\simeq3.7\,$eV. (c,d) EELS probability per unit of path length for $z_0=0.5\,$nm calculated using different models [see legend in (c)] for (c) different electron kinetic energies $E_0$ with fixed $N=5$ and (d) different $N$'s with $E_0=2.5\,$keV.}
\label{Fig3} 
\end{figure}

The loss function ${\rm Im}\{\rp\}$ provides a convenient way to represent the plasmon dispersion relation, as plasmons produce sharp features in the Fresnel reflection coefficient for p polarization. A weighted integral of this quantity over in-plane wave vectors gives the EELS probability under parallel aloof interaction [Fig.\ \ref{Fig3}(a)] according to Eq.\ (\ref{eq:EELS_par}). However, the integration limit has a threshold at $\omega=Qv$ and the weighting factor multiplying the loss function in the integrand diverges precisely at that point. The cutoff condition $\omega=Qv$ is represented in Fig.\ \ref{Fig3}(b) for different electron velocities (white lines) along with the loss function (density plot). As expected, the points of intersection with the plasmon band produce a dominant contribution that pops up as sharp peaks in the resulting EELS spectra [Fig.\ \ref{Fig3}(c,d)]. An increase in electron velocity (i.e., in the slope of the threshold line) results in a redshift of the spectral peak [Fig.\ \ref{Fig3}(c)], and likewise, thinner films show plasmons moving farther away from the $\omega=Qc$ light line, thus producing shifts toward higher plasmon energies in the EELS spectra for fixed electron energy. We remark that RPA and classical calculations lead to quantitatively similar results for this configuration, and the former are roughly independent of the choice of confining electron potential.

\begin{figure}
\centering
\includegraphics[width=0.35\textwidth]{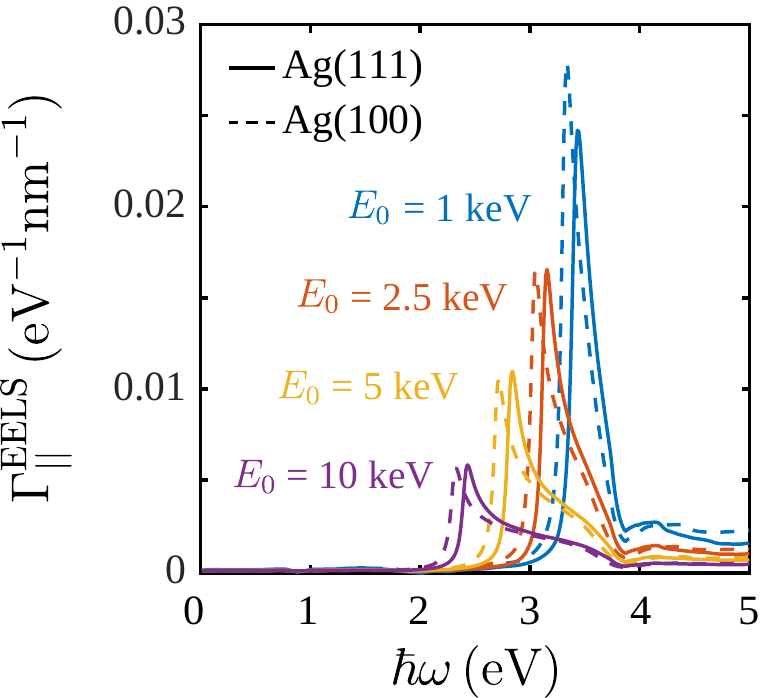}
\caption{Plasmon dependence on crystallographic surface orientation: Ag(111) and Ag(100) films. We compare EELS spectra calculated in the ALP model under the same conditions as in Fig.\ \ref{Fig3}(c,d) for $N=13\,$ML Ag(111) and $N=15\,$ML Ag(100) films (thickness ratio differing by $<0.1\%$).}
\label{Fig4} 
\end{figure}

The ALP model incorporates experimental information on electronic bands, which depend on crystallographic orientation (see Table\ \ref{Table1}). We explore the effects of this dependence by comparing aloof EELS spectra obtained from Ag(111) and Ag(100) films in Fig.\ \ref{Fig4}. In order to eliminate discrepancies arising from differences in thickness, we consider films consisting of $N=13$ and $N=15$\,MLs, respectively, so that the thickness ratio is $(2/\sqrt{3})\times(13/15)\approx1.001$. We remind that Ag(111) displays a projected bulk gap in the electronic bands, in contrast to Ag(100) [see Fig.\ \ref{FigS1}(a) in Appendix\ \ref{AFSI}]; as a consequence the former supports electronic surface states unlike the latter \cite{CSE99}. Despite these remarkable differences in electronic structure, the resulting spectra look rather similar, except for a small redshift of Ag(100) plasmon peaks relative to Ag(111), comparable in magnitude to those observed in semi-infinite Ag(111) and Ag(110) crystal surfaces through angle-resolved low-energy EELS \cite{STP1989}, although the actual magnitude of the shift might be also influenced by electron confinement in our ultrathin films.

\begin{figure}
\centering
\includegraphics[width=0.35\textwidth]{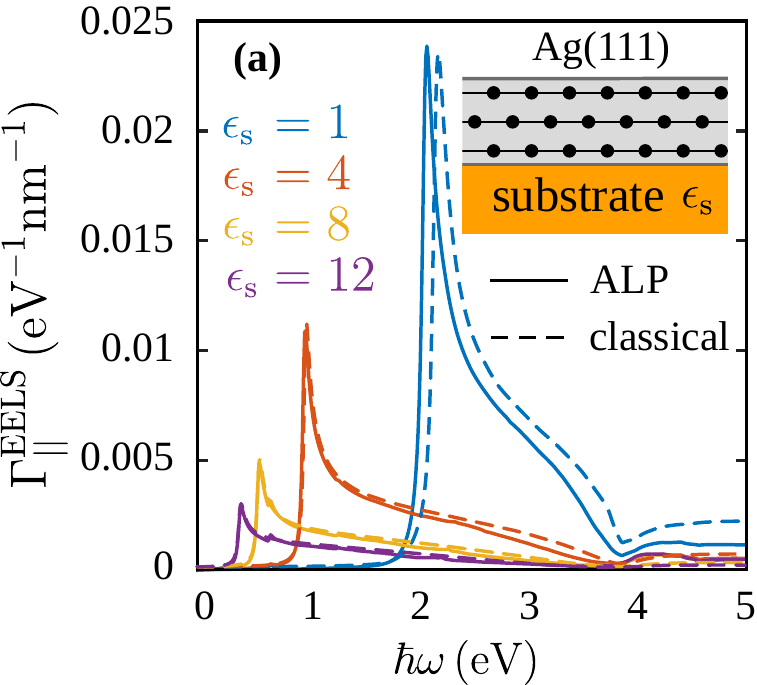}
\caption{Substrate-induced plasmon shift. We show EELS spectra for 2.5\,keV electrons calculated in either the ALP model or the local classical description under the same conditions as in Fig.\ \ref{Fig3} for a Ag(111) film consisting of $N=5$\,MLs supported on a planar dielectric substrate of permittivity $\epsilon_{\rm s}$ as indicated by labels.}
\label{Fig5}
\end{figure}

The presence of a dielectric substrate of permittivity $\epsilon_{\rm s}$ is known to redshift the plasmon frequency of thin films by a factor $\sim1/\sqrt{1+\epsilon_{\rm s}}$ due to the attractive image interaction \cite{paper235}. This effect is observed in our calculated aloof EELS spectra, for which we obtain the combined film-substrate reflection coefficient by using a Fabry-Perot approach, as discussed elsewhere \cite{paper329}. We find again excellent agreement between RPA simulations using the ALP potential and classical calculations [Fig.\ \ref{Fig5}], and in fact, the resemblance between the spectral profiles obtained with both methods increases with $\epsilon_{\rm s}$.

\begin{figure}
\centering
\includegraphics[width=0.45\textwidth]{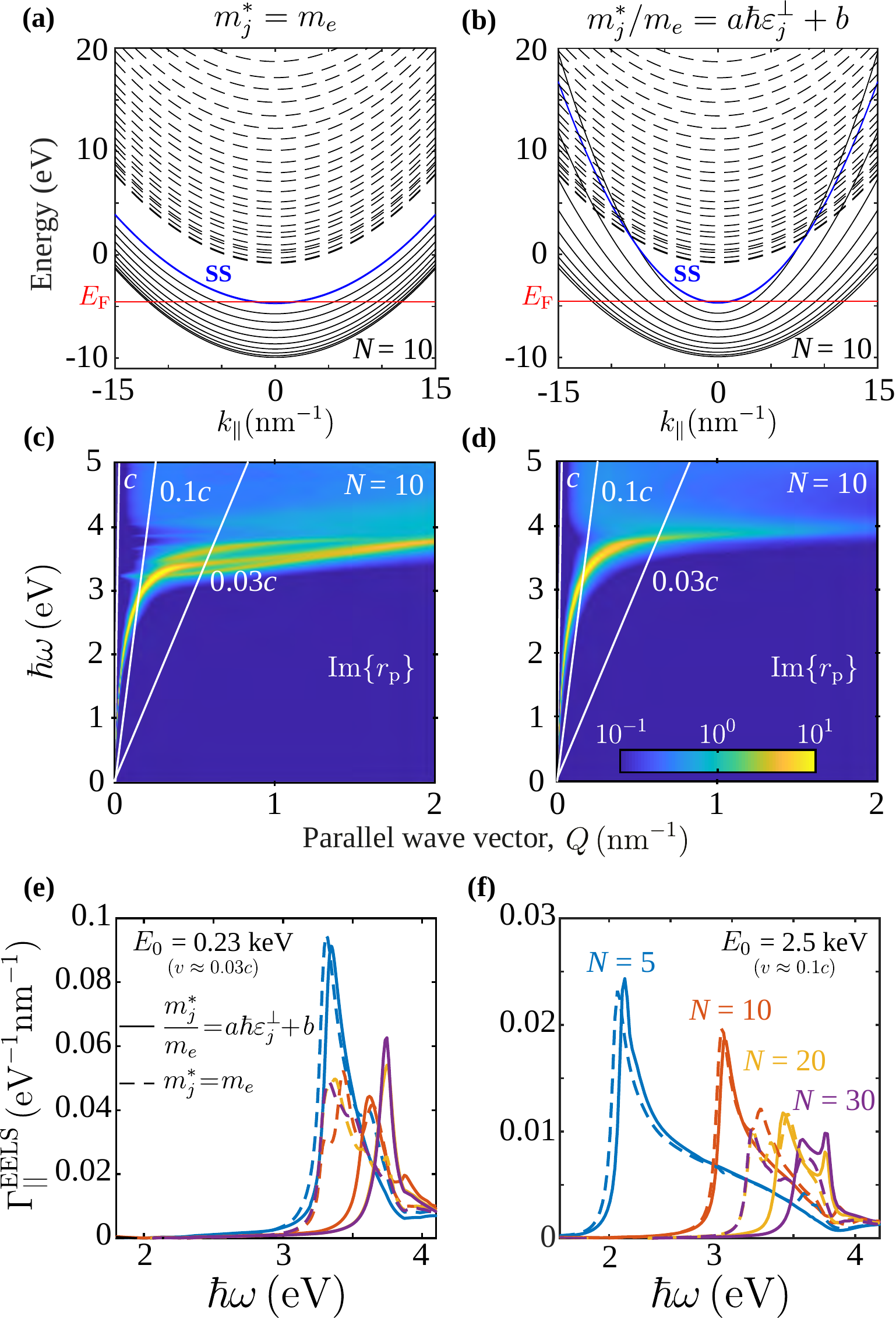}
\caption{The role of the electron effective mass. (a,b) In-plane parabolic QW bands of a $N=10\,$ML Ag(111) film in the ALP model with (a) constant and (b) energy-dependent effective mass ($m_j^*=\me$ and $m_j^*=(a\hbar\en_j^\perp+b)\me$, respectively, see Table\ \ref{Table1}). The surface state bands (blue curves) have a mass $0.4\,\me$. Solid (dashed) curves represent bands that are occupied (unoccupied) at $\kpar=0$. The Fermi level is shown as a horizontal red line. (c,d) Loss function ${\rm Im}\{\rp\}$ under the conditions of (a,b), respectively. (e,f) EELS probability under parallel aloof interaction at a distance $z_0=0.5\,$nm for two different electron energies corresponding to the $\omega=Qv$ lines shown in (c,d) and different film thicknesses (see labels) calculated in the ALP model with constant (dashed curves) and energy-dependent (solid curves) electron effective mass.}
\label{Fig6} 
\end{figure}

In Fig.\ \ref{Fig6} we examine the way lateral dispersion of QW states affects the plasmonic properties of ultrathin Ag films when using the ALP potential. Comparison of the band structures calculated with [Fig.\ \ref{Fig6}(b)] and without [Fig.\ \ref{Fig6}(a)] inclusion of an energy dependence in the in-plane effective mass anticipates a clear difference between the two of them: the latter shows the same energy jumps between different bands irrespective of the electron parallel wave vector $\kpar$; those energy jumps will therefore be favored in the optical response, giving rise to spurious spectral features. In contrast, differences in lateral dispersion associated with the energy dependence of the effective mass (described here by fitting existing angle-resolved photoemission data \cite{RNS01,SPC05,GPC03,MMC90,MTH04}) should at least partially wash out those spectral features. This is clearly observed in the resulting dispersion diagrams [Fig.\ \ref{Fig6}(c,d)] and aloof EELS spectra [Fig.\ \ref{Fig6}(e,f)]. In particular, the dispersion relation for constant $m^*_j$ [Fig.\ \ref{Fig6}(c)] reveals a complex mixture of resonances at energies above 3\,eV, which we find to be strongly affected by the HOMO-LUMU gap energy (not shown); these resonances cause fine structure in the EELS spectra that disappears when a realistic energy dependence is introduced in the lateral effective mass [Fig.\ \ref{Fig6}(e)].

\begin{figure}
\centering
\includegraphics[width=0.45\textwidth]{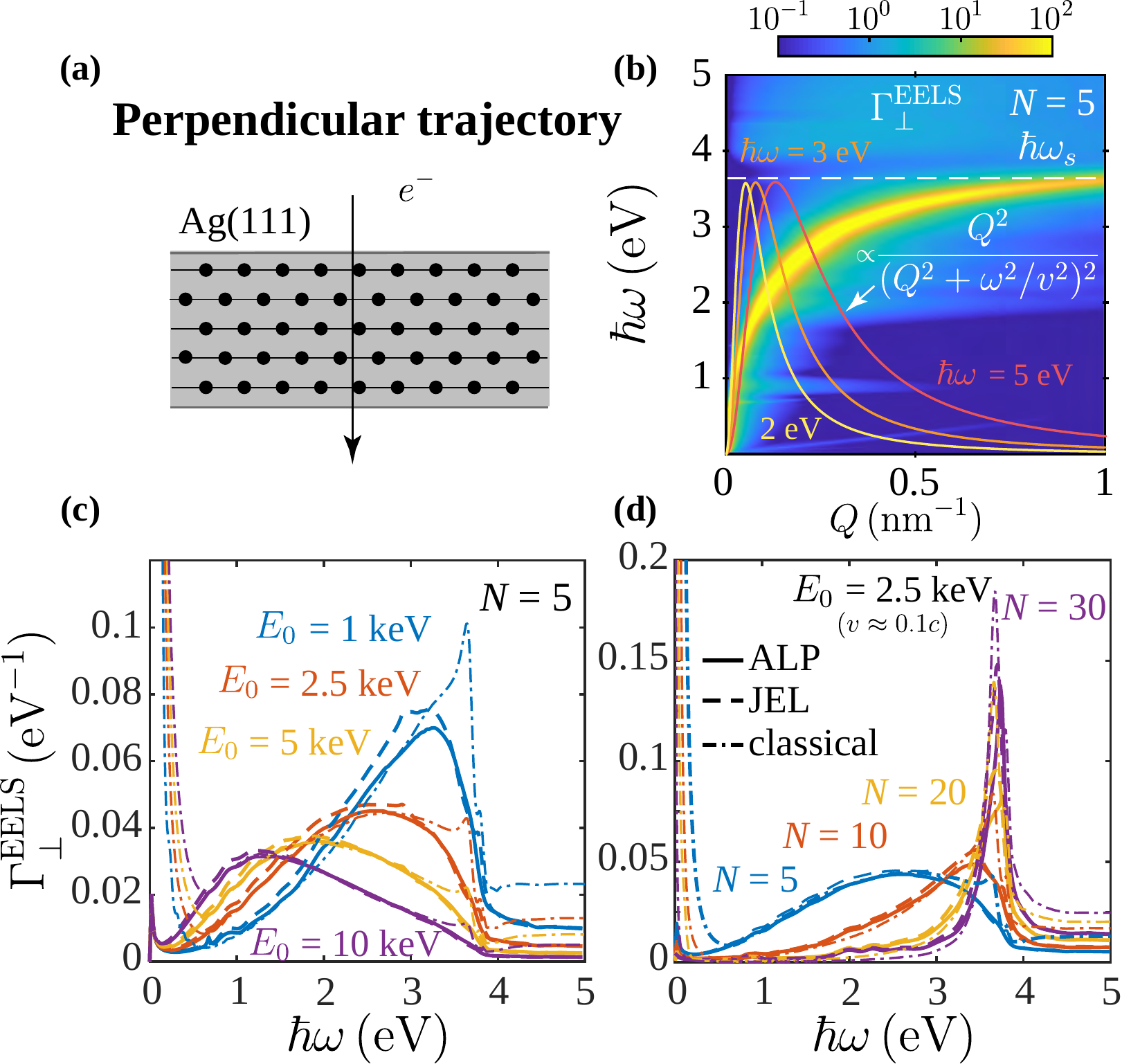}
\caption{EELS in thin Ag(111) films under normal incidence. (a) Scheme showing an electron normally traversing a $N=5$\,ML Ag(111) metal film. (b) Momentum- and energy-resolved EELS probability $\Gamma_{\perp}^\EELS(Q,\omega)$ [Eq.\ (\ref{eq:EELS_perp_kernel})] calculated for $E_0=2.5\,$keV electrons ($v/c\approx0.1$) in the ALP model for the film shown in (a). Colored solid curves show $Q^2/(Q^2+\omega^2/v^2)^2$ profiles as a function of $Q$ for different energy losses $\hbar\omega=2$, 3, and 5\,eV, while the dashed horizontal line indicates $\hbar\ws$. (c,d) EELS probability calculated using different models [see legend in (d)] for (c) different electron kinetic energies $E_0$ with fixed $N=5$ and (d) different $N$'s with $E_0=2.5\,$keV.}
\label{Fig7} 
\end{figure}

We also analyse EELS spectra for normally impinging electron beams [Fig.\ \ref{Fig7}]. The momentum- and energy-resolved EELS probability given by Eq. \eqref{eq:EELS_perp_kernel} reveals the plasmon dispersion in analogy to the loss function [cf. Figs.\ \ref{Fig3}(b) and \ref{Fig7}(b)]. But now, this quantity is directly accessible under normal incidence by recording angle- and energy-dependent electron transmission intensities, as already done in pioneering experiments for thicker Al films showing both bonding and antibonding plasmon dispersions \cite{VS1973}. In contrast to the aloof configuration, the transmission EELS spectra exhibit broader plasmon features [Fig.\ \ref{Fig7}(c,d)], which in the thin film limit \cite{paper228} are the result of weighting the loss function with a profile $Q^2/(Q^2+\omega^2/v^2)^2$ [see also Eq.\ (\ref{aaaa}), where an extra factor of $Q$ emerges from $\chi$ in the small $Q$ limit], represented in Fig.\ \ref{Fig7}(b) for 2.5\,keV electrons and different energies $\hbar\omega$ (colored curves); these spectra reveal indeed a broad spectral overlap with the plasmon band. Again, we observe very similar results from RPA and classical descriptions, and just a minor dependence on electron potential in the former.

\begin{figure}
\centering
\includegraphics[width=0.45\textwidth]{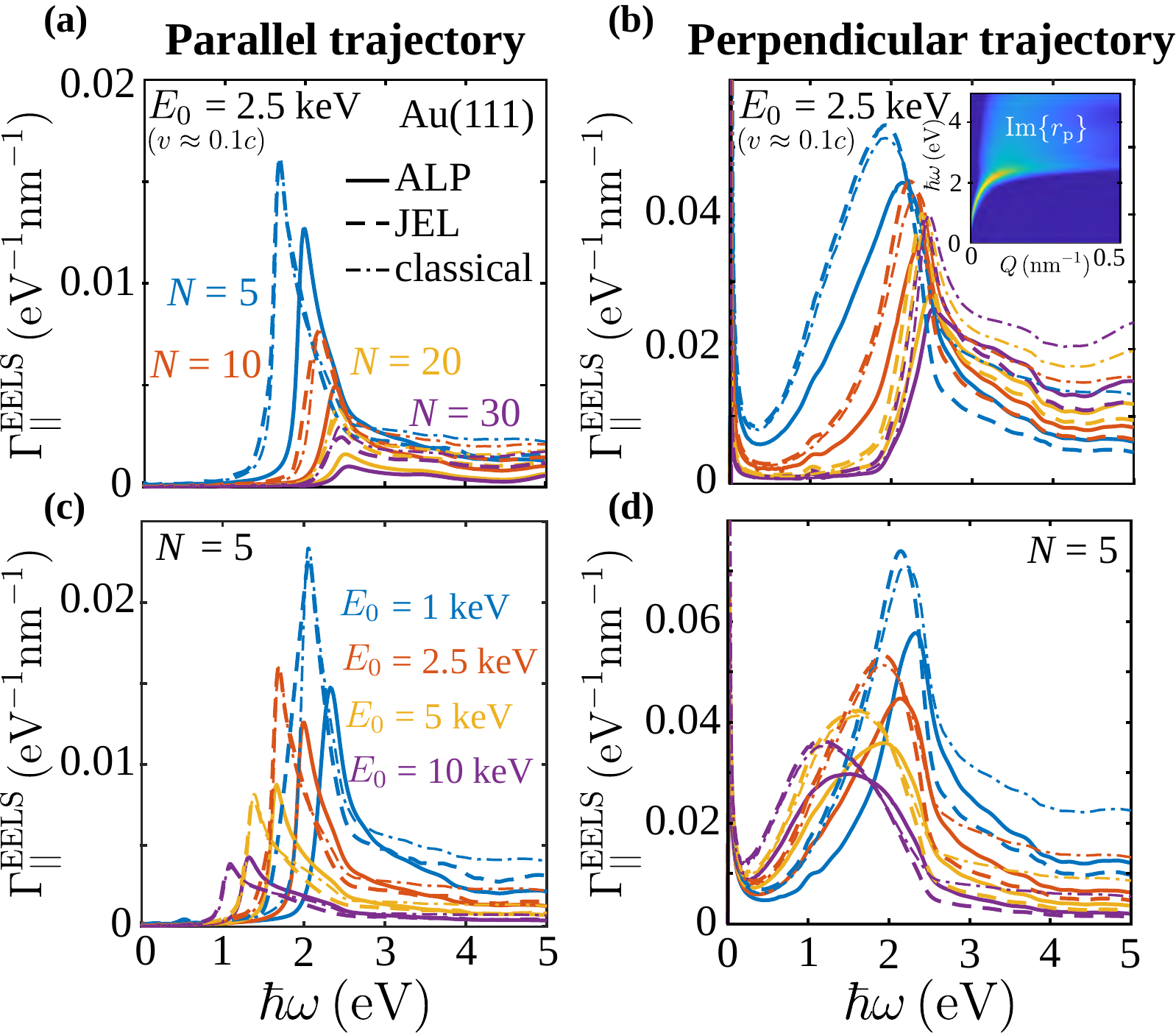}
\caption{EELS spectra for gold Au(111) films. We consider for  (a,c) aloof and (b,d) normal trajectories for either (a,b) fixed electron energy ($E_0=2.5\,$keV) and varying film thickness ($N=5$-30\,MLs) or (c,d) fixed $N=5$ and varying electron energy. Calculations for the same models as in Fig.\ \ref{Fig3} are presented. The plasmon dispersion is shown for $N=10$\,MLs using the ALP model in the inset of (b).}
\label{Fig8}
\end{figure}

We conclude by showing EELS calculations for Au(111) films in Fig.\ \ref{Fig8}. This noble metal has a similar conduction electron density as Ag, but the Au d-band is closer to the Fermi energy, therefore producing large screening ($\epsilon_{\rm b}\sim9$ in the plasmonic region) compared with Ag ($\epsilon_{\rm b}\sim4$, see Fig.\ S4 in SI). This causes a shift of the high-$Q$ surface plasmon asymptote down to $\hbar\ws\simeq2.5\,$eV. Additionally, damping is also stronger (more than three times larger than in Ag, see Appendix\ \ref{bsi}), which results in broader spectral features [cf. Fig.\ \ref{Fig8} for Au and Figs.\ \ref{Fig3}(c,d) and \ref{Fig7}(c,d)]. Interestingly, we observe significant blue shifts in the plasmon spectral features when using the ALP potential as compared with both jellium DFT and classical models. This effect could originate in a more substantial role played by the electronic band structure in Au(111) because the projected bulk gap extends further below the Fermi level, and additionally, the surface state band is also more deeply bound [see Fig.\ \ref{FigS1}(b) in Appendix\ \ref{AFSI}]. This is consistent with the general dependence of the optical surface conductivity on Fermi momentum $\kF$ and velocity $\vF$: in the Drude model for graphene and the two-dimensional electron gas, this quantity is proportional to $\kF\vF$ and the surface plasmon frequency scales as $\propto\sqrt{\kF\vF}$; the situation is more complicated in our thin films because they have multiple 2D bands crossing the Fermi level, but the presence of a deeper gap in Au(111) indicates that the effective band-averaged value of $\kF\vF$ (i.e., with $\kF$ defined by the crossing of each QW at the Fermi level and $\vF$ as the slope of the parabolic dispersion at that energy) is larger than in Ag surfaces, characterized by the presence of shallower bands near $\EF$; we thus expect an increase in Drude weight, and consequently, a plasmon blue shift, in Au(111) relative to Ag; this argument is reinforced by the small effective mass of surface states in Au(111) compared with Ag(111), which also pushes up their associated $\vF$. In summary, the plasmon blue shifts observed in Au(111) when using the realistic ALP potential seem to have a physical origin, although more sophisticated first principles simulations might be needed to conclusively support this finding.

% ----------------------------------------------------------------------------
\section{Conclusion}

In summary, we have shown that a local classical dielectric model predicts reasonably well the intensities and dispersion relations of plasmons in ultrathin silver films when compared to quantum-mechanical simulations based on the RPA with different potentials used to simulate the conduction one-electron wave functions. We attribute the small effect of nonlocality in the plasmonic response of these films to the fact that their associated electron motion takes place along in-plane directions, in contrast to metal nanoparticles with a similar size as the film thickness here considered (i.e., electron surface scattering is unavoidable in such particles, thus introducing important nonlocal effects). We confirm this agreement between classical and quantum simulations in Ag films down to a few atomic layers in thickness \cite{paper329,SSP19}, consistent with previous smooth-interface hydrodynamic theory \cite{paper244}. Additionally, our quantum RPA simulations are relatively insensitive to the details of the confining electron potential, so similar results are obtained when using either a smooth jellium DFT model or a phenomenological potential that incorporates atomic-layer corrugation to fit relevant elements of the electronic band structure. In particular, the latter produces results that are rather independent of the crystallographic orientation of the film. Nonetheless, it is important to introduce the correct energy dependence of the out-of-plane effective mass in the phenomenological potential model, as otherwise spurious features show up in the calculated plasmon spectra. Although these potentials lead to substantially different plasmon charge distributions, spatial integration gives rise to similar plasmon dispersion relations. Interestingly, band effects described in the ALP potential model are more significant in Au, where they produce plasmon blue shifts relative to the predictions of classical and jellium DFT simulations; we attribute this different behavior in Au(111) relative to Ag(111) and Ag(100) to the fact that the former surface exhibits a projected bulk gap that extends further below the Fermi level, and additionally, this gives rise to more bound surface states. We remark that EELS provides the means to access the dispersion relations of strongly confined plasmons in ultrathin metal films, which are too far from the light line to be measured by means of optical techniques.

\acknowledgements
This work has been supported in part by the ERC (Advanced Grant 789104-eNANO), the Spanish MINECO (MAT2017-88492-R and SEV2015-0522), the Catalan CERCA Program, the Fundaci\'o Privada Cellex, and the Quscope center sponsored by the Villum Foundation.

\appendix

\section{Background screened interaction}
\label{bsi}

We introduce the effect of interband polarization in the plasmonic spectral region of noble metals through a dielectric slab of permittivity $\epsb(\omega)=\epsm(\omega)+\wp^2/\omega(\omega+\ii\gamma^{\rm exp})$, that is, the local dielectric function of the bulk metal $\epsm(\omega)$ from which we subtract a classical bulk Drude term representing the contribution of conduction electrons. In practice, we take $\epsm(\omega)$ from measured optical data \cite{JC1972} and use parameters $\hbar\wp=9.17\,$eV and $\hbar\gamma^{\rm exp}=21\,$meV for Ag, and $\hbar\wp=9.06\,$eV and $\hbar\gamma^{\rm exp}=71\,$meV for Au. The resulting $\epsb(\omega)$ is plotted in Fig.\ \ref{FigS4} in Appendix\ \ref{AFSI}. Incidentally, as we explain in Sec.\ \ref{Rso}, we set the damping parameter to $\gamma=\gamma^{\rm exp}/2$ in the RPA formalism in order to fit the experimental plasmon width. Following previous work \cite{L93}, we take the background dielectric slab to have a thickness $\dm=Na_s$, where $N$ is the number of atomic layers and $a_s$ is the interlayer spacing, so that it extends symmetrically a distance $a_s/2$ outside the outer atomic plane on each side of the film.

We reproduce for convenience a previously reported expression \cite{paper329} for the screened interaction, used here to account for background polarization in the a self-standing metal film of thickness $\dm$ and background permittivity $\epsb$ contained in the $0<z<d$ region:
\begin{align}
\uupsilon(Q,z,z')=\uupsilon^{\rm dir}(Q,z,z')+\uupsilon^{\rm ref}(Q,z,z'),
\nonumber
\end{align}
where
\begin{align}
\uupsilon^{\rm dir}(Q,z,z')=\frac{2\pi}{Q}\,\ee^{-Q|z-z'|}\times
\left\{\begin{array}{ll}
1,  &z,z'\le0 \text{ or } z,z'>\dm \\
1,  &0<z,z'\le\dm \\
0, & \text{otherwise}
\end{array} \right.
\nonumber
\end{align}
and
\begin{widetext}
\begin{align}
\uupsilon^{\rm ref}(Q,z,z')
=\frac{(2\pi/Q)}{(\epsb+1)^2-(\epsb-1)^2\ee^{-2Q\dm}} \times
\left\{\begin{array}{ll}
(1-\epsb^2)\left(\ee^{2Q\dm}-1\right)\,\ee^{-Q(z+z')},  &\dm<z, z' \\ %-----
2\left[(\epsb+1)\ee^{-Q(z-z')}+(\epsb-1)\ee^{-Q(z+z')}\right],  & 0<z'\le\dm<z \\ %-----
4\epsb\;\ee^{-Q(z-z')},  & z'\le0\;\;\text{and}\;\;\dm<z \\ %-----
2\left[(\epsb+1)\ee^{Q(z-z')}+(\epsb-1)\ee^{-Q(z+z')}\right],  & 0<z\le\dm<z' \\ %-----
(1/\epsb)\big\{
(\epsb^2-1)\left[\ee^{-Q(z+z')}+\ee^{-Q(2\dm-z-z')}\right] & \nonumber\\
\quad\;\;\;
+(\epsb-1)^2\left[\ee^{-Q(2\dm+z-z')}+\ee^{-Q(2\dm-z+z')}\right]
\big\},  & 0<z,z'\le\dm \\ %-----
2\left[(\epsb+1)\ee^{-Q(z-z')}+(\epsb-1)\ee^{-Q(2\dm-z-z')}\right],  & z'\le0<z\le\dm \\ %-----
4\epsb\;\ee^{Q(z-z')},  &z\le0\;\;\text{and}\;\;\dm<z'\\ %-----
2\left[(\epsb+1)\ee^{Q(z-z')}
+(\epsb-1)\ee^{-Q(2\dm-z-z')}\right],  &z\le0<z'\le\dm\\ %-----
(1-\epsb^2)\left(1-\ee^{-2Q\dm}\right)\,\ee^{Q(z+z')}.  &z, z'\le0
\end{array} \right.
\nonumber
\end{align}
\end{widetext}

For completeness, we illustrate the dramatic effects of interband processes in Fig.\ \ref{FigS5} in Appendix\ \ref{AFSI} by comparing calculations obtained for Ag films using either screened or bare Coulomb interactions.

\begin{widetext}

\section{Additional figures}
\label{AFSI}

\clearpage

\begin{figure}
\centering
\includegraphics[width=0.60\textwidth]{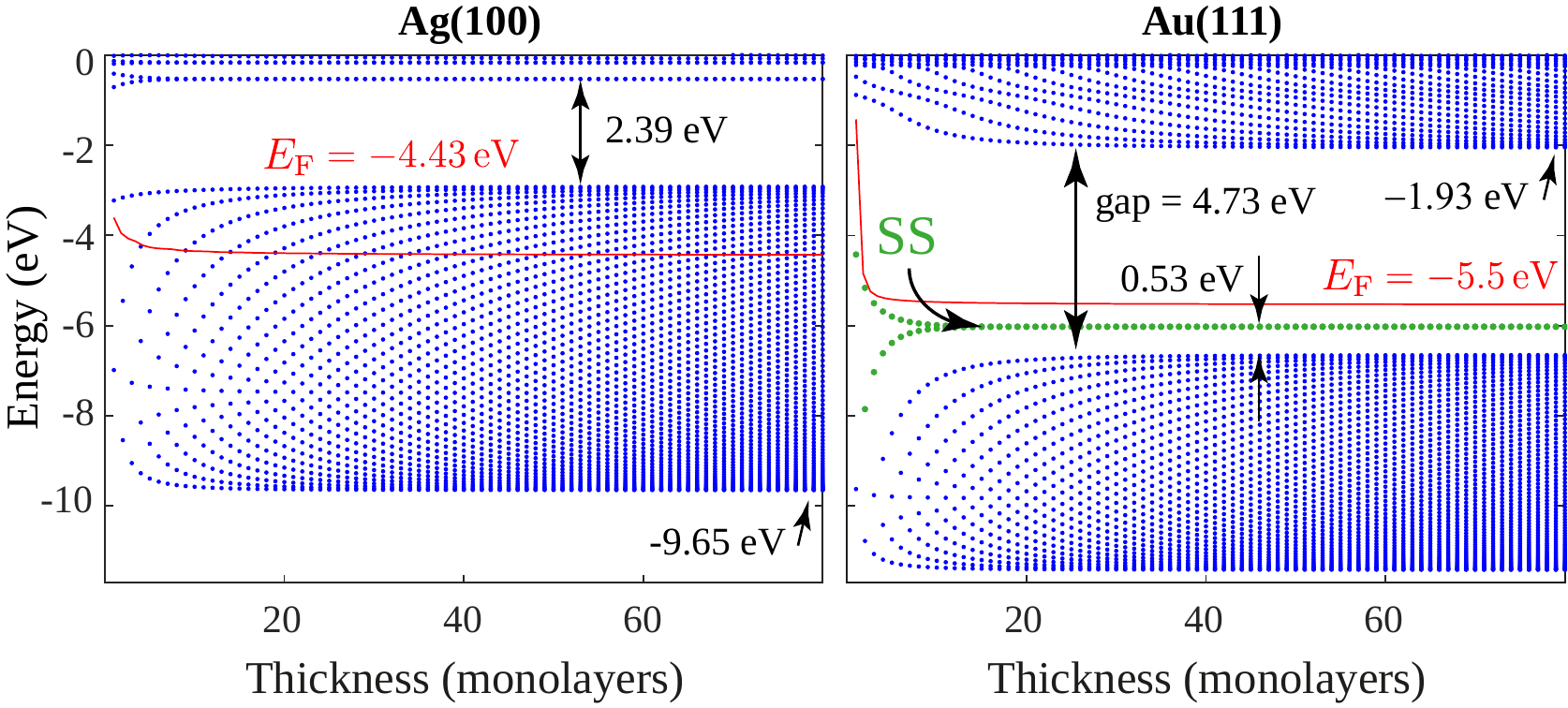}
\caption{ALP model calculations similar to those of Fig.\ 1(d), but for Ag(100) and Au(111) films.}
\label{FigS1}
\end{figure}

\begin{figure}
\centering
\includegraphics[width=0.60\textwidth]{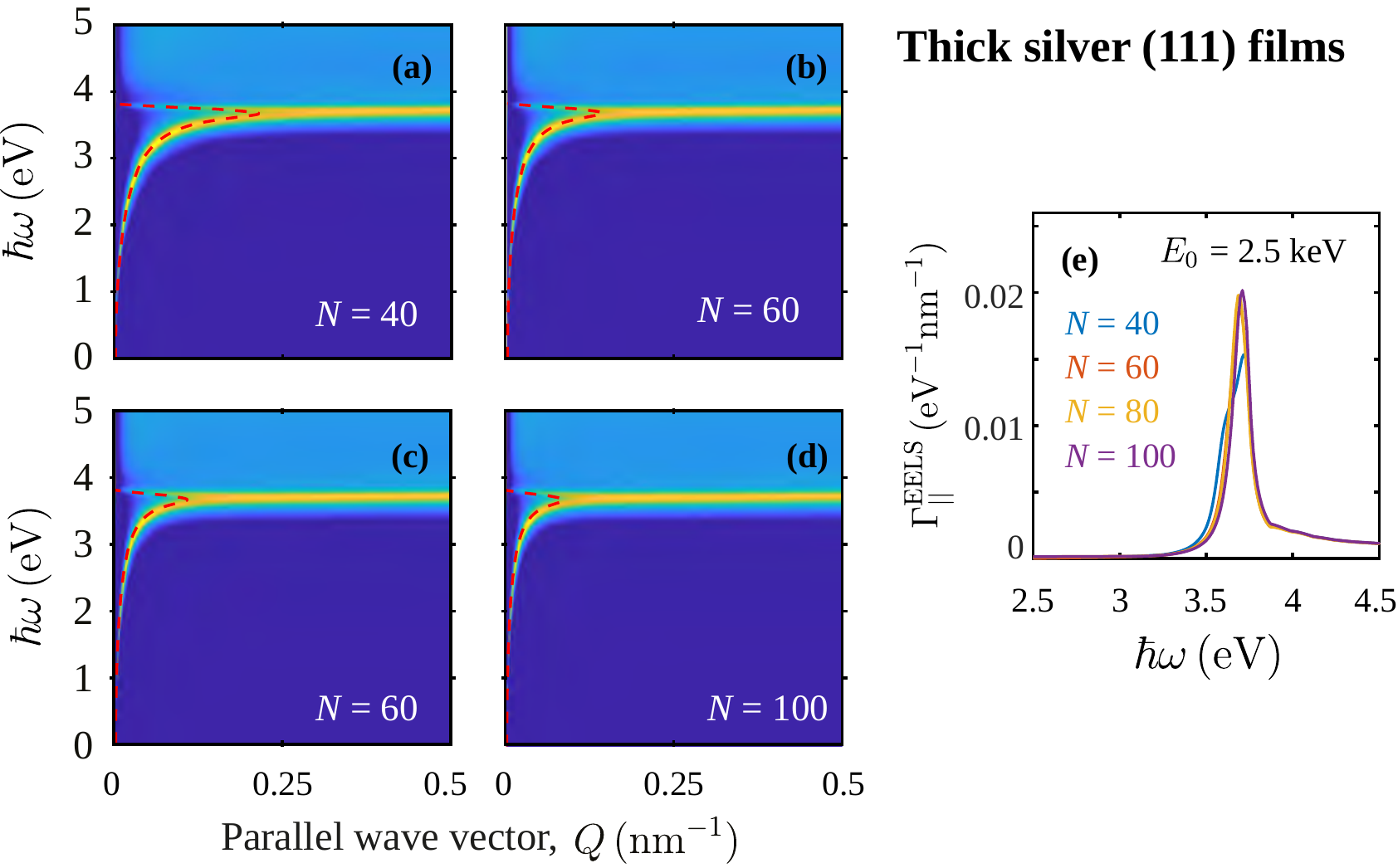}
\caption{ALP model calculations for (a-d) the loss function ${\rm Im}\{\rp\}$ of Ag(111) films of different thickness $N$ and (b) the resulting EELS spectra of normally incident 2.5\,keV electrons.}
\label{FigS2}
\end{figure}

\begin{figure}
\centering
\includegraphics[width=0.60\textwidth]{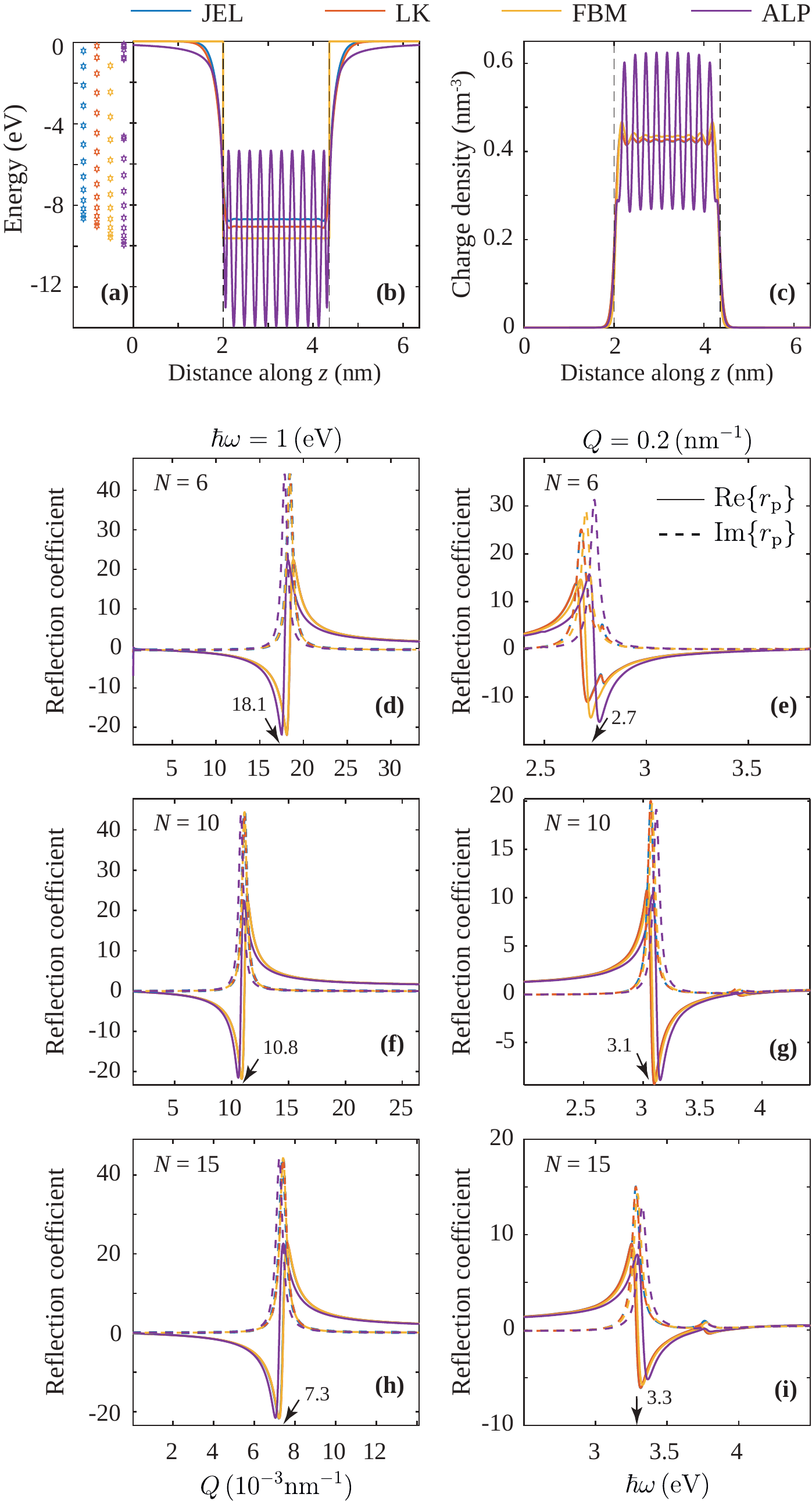}
\caption{Dependence of the RPA response on model potential. We show (a) the binding conduction electron energies, (b) the confining potential, and (c) the conduction electron density for a $N=10$\,ML Ag(111) film, as well as (d-i) the reflection coefficient $\rp$ of Ag(111) films of different thickness $N$ for either (d,f,h) fixed photon energy $\hbar\omega$ as a function parallel wave vector $Q$ or (e,g,i) fixed $Q$ as a function of $\hbar\omega$. We calculate $\rp$ in the RPA and consider different confining electron potentials, as indicated by the upper labels: JEL and ALP, defined in the main text; LK, a superposition of the parametrized jellium DFT potential for semi-infinite surfaces taken from Lang and Kohn [Phys. Rev. B {\bf 1}, 4555 (1970)] for a one-electron radius $r_s=3\,$a.u., adopted for each of the film surfaces and glued by hand at the film center; and FBM, a square-well finite-barrier model potential. Only the ALP incorporates an energy dependence on the lateral effective mass, while the rest of the models assume $m^*_j=\me$.}
\label{FigS3}
\end{figure}

\begin{figure}
\centering
\includegraphics[width=0.55\textwidth]{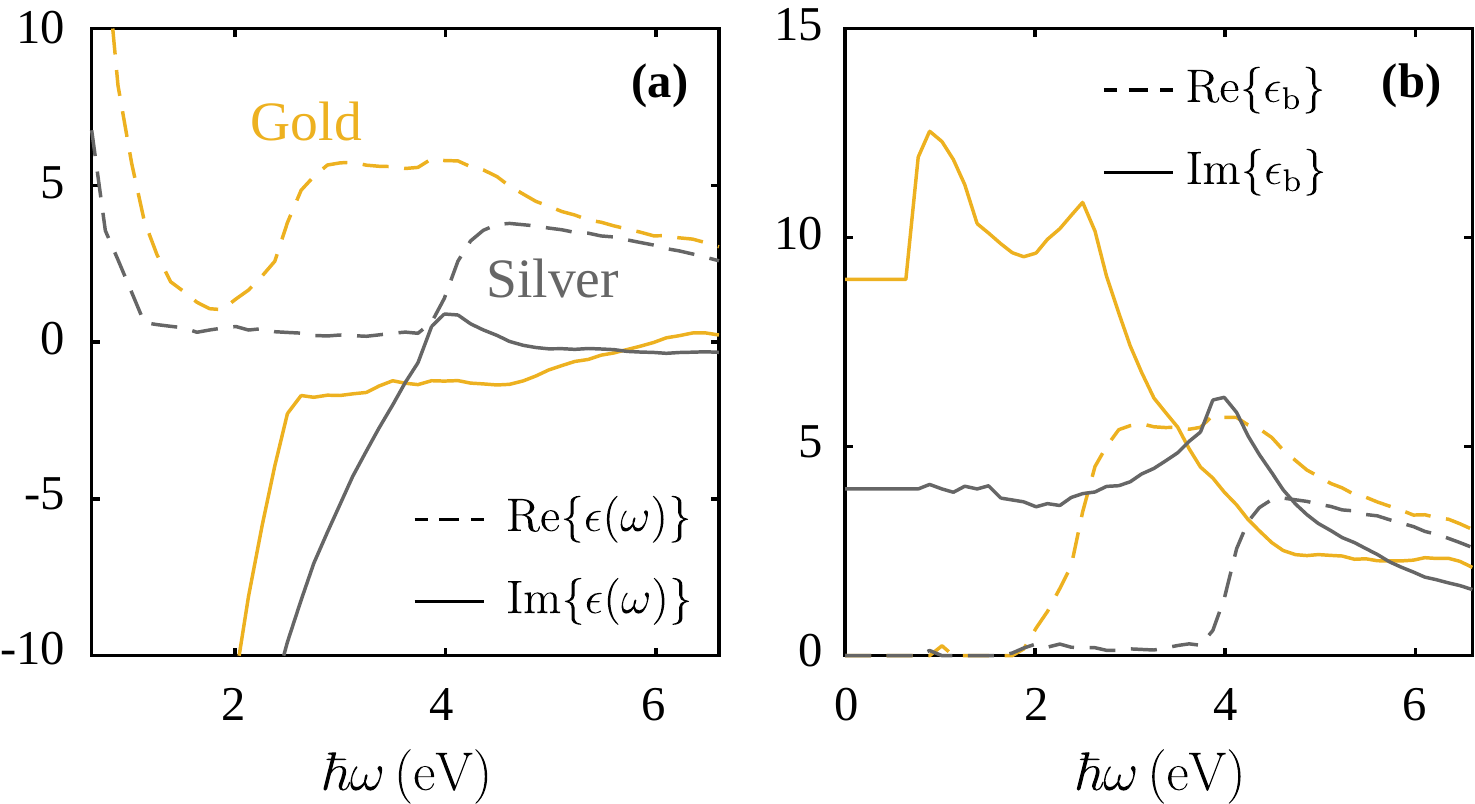}
\caption{Dielectric function $\epsilon(\omega)$ of Ag and Au taken from Jonhson and Christy [Phys. Rev. B {\bf 6}, 4370 (1972)] and background permittivity $\epsilon_{\rm b}(\omega)=\epsilon(\omega)+\wp^2/\omega(\omega+\ii\gamma^{\rm exp})$ obtained by subtracting a Drude term with parameters $\hbar\wp=9.17\,$eV and $\hbar\gamma^{\rm exp}=21\,$meV for Ag, and $\hbar\wp=9.06\,$eV and $\hbar\gamma^{\rm exp}=71\,$meV for Au. We take $\epsilon_{\rm b}=4$ and $\epsilon_{\rm b}=9.5$ for Ag and Au in the $\hbar\omega<0.6\,$eV region, which is not covered in the above reference.}
\label{FigS4}
\end{figure}

\begin{figure}
\centering
\includegraphics[width=0.55\textwidth]{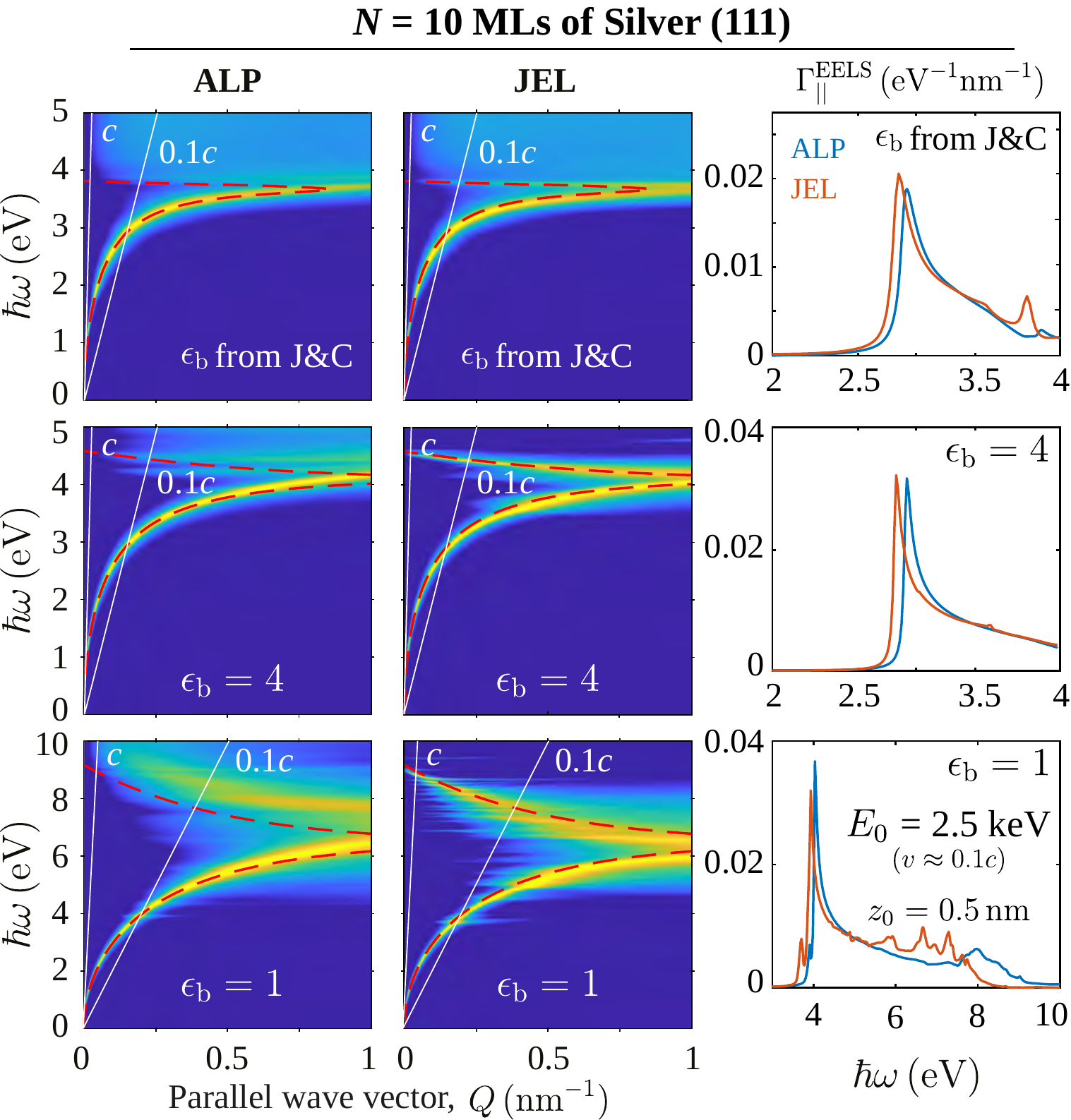}
\caption{Effect of background screening. We show dispersion diagrams (color plots) showing the loss function ${\rm Im}\{\rp\}$ of $N=10$\,ML Ag(111) films as calculated in the RPA for ALP (left) and JEL (center) model potentials when the background permittivity $\epsilon_{\rm b}$ is obtained from optical data (upper plots, see Fig.\ \ref{FigS4}) or set to a constant value $\epsilon_{\rm b}=4$ (middle) or $\epsilon_{\rm b}=1$ (bottom). We find $\epsilon_{\rm b}=4$ to represent approximately well background screening in silver over the plasmonic spectral region, whereas $\epsilon_{\rm b}=1$ gives rise to unrealistic blue shifts. These conclusions are maintained when examining aloof EELS spectra (right plots, calculated for 2.5\,keV electrons passing at a distance of 0.5\,nm from the metal surface).}
\label{FigS5}
\end{figure}

\end{widetext}

\clearpage

%\bibliographystyle{apsrev}
%\bibliography{../../../bibtex/refs}

\end{document}